\documentclass[runningheads]{llncs}
\usepackage{graphicx}

\usepackage{multirow}
\usepackage{subcaption}
\usepackage{hyperref}
\usepackage{float}
\usepackage{algorithm}
\usepackage{algpseudocode}

\usepackage{amsfonts}
\usepackage{amsmath}
\usepackage{wrapfig}

\begin{document}
\title{Process Variant Analysis Across Continuous Features: A Novel Framework\thanks{This research was supported by the research training group ``Dataninja'' (Trustworthy AI for Seamless Problem Solving: Next Generation Intelligence Joins Robust Data Analysis) funded by the German federal state of North Rhine-Westphalia.}}
\titlerunning{Process Variant Analysis Across Continuous Features}
% If the paper title is too long for the running head, you can set
% an abbreviated paper title here
%
\author{Ali Norouzifar \inst{1}\orcidID{0000-0002-1929-9992} \and
Majid Rafiei\inst{1}\orcidID{0000-0001-7161-6927} \and
Marcus Dees\inst{2}\orcidID{0000-0002-6555-320X} \and
Wil van der Aalst\inst{1}\orcidID{0000-0002-0955-6940}}
\authorrunning{A. Norouzifar et al.}
% First names are abbreviated in the running head.
% If there are more than two authors, 'et al.' is used.
%
\institute{RWTH University, Aachen, Germany \\ \email{\{ali.norouzifar, majid.rafiei, wvdaalst\}@pads.rwth-aachen.de} \and
UWV Employee Insurance Agency, Amsterdam, Netherlands \\
\email{Marcus.Dees@uwv.nl}}
\maketitle              % typeset the header of the contribution
\setcounter{footnote}{0}

\begin{abstract}
Extracted event data from information systems often contain a variety of process executions making the data complex and difficult to comprehend. Unlike current research which only identifies the variability over time, we focus on other dimensions that may play a role in the performance of the process. This research addresses the challenge of effectively segmenting cases within operational processes based on continuous features, such as duration of cases, and evaluated risk score of cases, which are often overlooked in traditional process analysis. We present a novel approach employing a sliding window technique combined with the earth mover's distance to detect changes in control flow behavior over continuous dimensions. This approach enables case segmentation, hierarchical merging of similar segments, and pairwise comparison of them, providing a comprehensive perspective on process behavior. We validate our methodology through a real-life case study in collaboration with UWV, the Dutch employee insurance agency, demonstrating its practical applicability. 
This research contributes to the field by aiding organizations in improving process efficiency, pinpointing abnormal behaviors, and providing valuable inputs for process comparison, and outcome prediction.

\end{abstract}

\vspace{-30pt}
\section{Introduction}
Process mining techniques are used to analyze event data generated by different types of information systems. For instance, performance analysis using process mining techniques has provided a range of new opportunities for business owners to analyze and improve their processes. From the performance point of view, we often observe that the execution policies may vary significantly for different groups of process instances based on their characteristics.

\begin{figure}[tb]
\begin{subfigure}{.95\textwidth}
\centering
\includegraphics[scale=0.4]{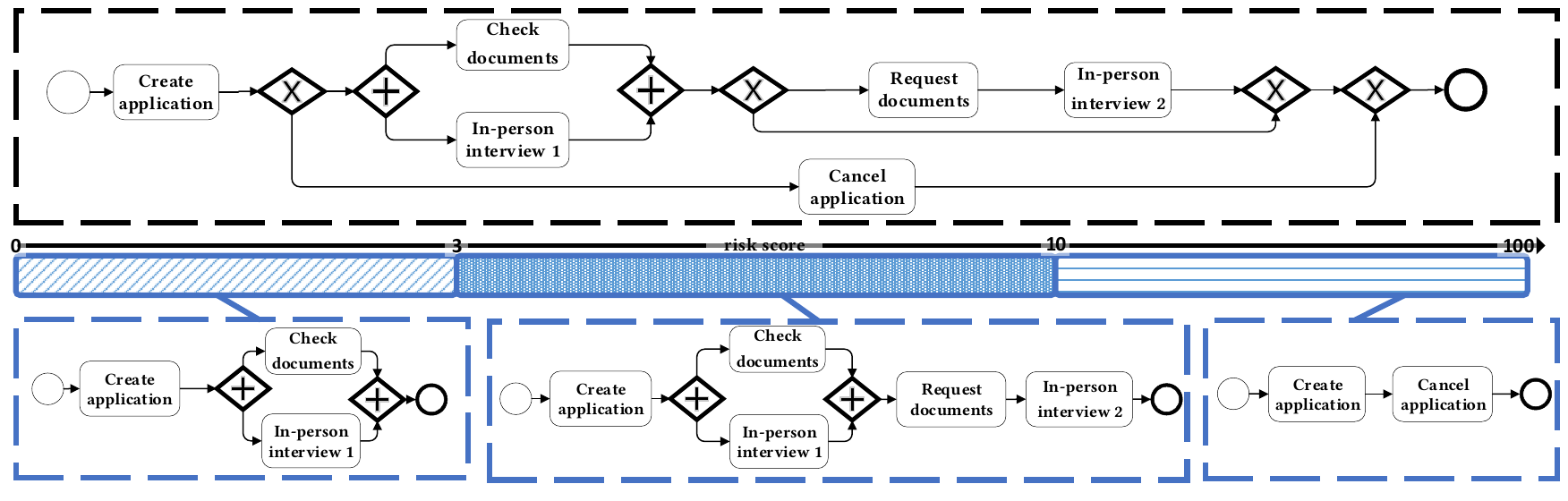}
\caption{The BPMN on top shows a claim handling process in which each case has a risk score and may have a different handling procedure based on the risk score.}
\label{intro_example}
\end{subfigure}\\
\begin{subfigure}{.95\textwidth}
\centering
\includegraphics[scale=0.25]{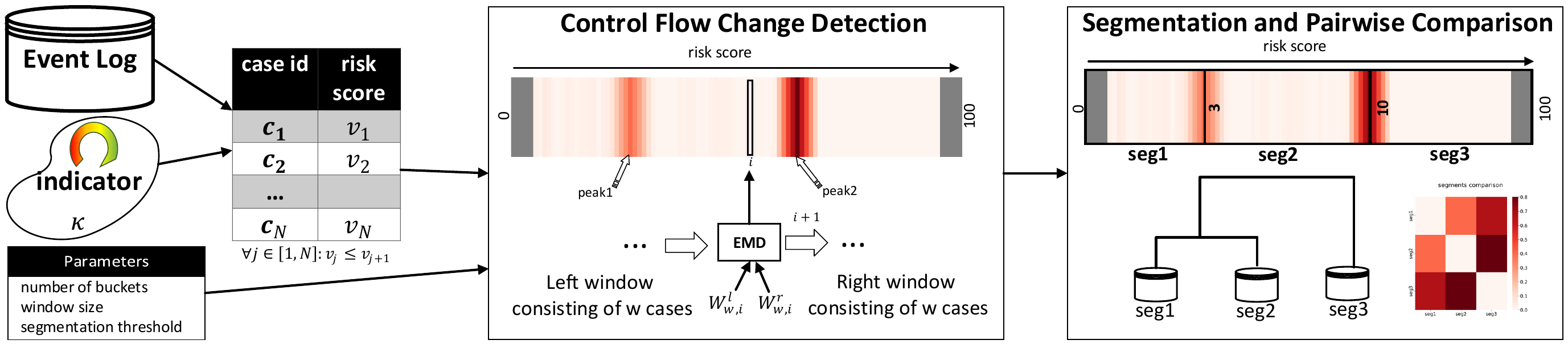}
\caption{The proposed framework for process variant analysis based on the risk score.}
\label{ov_new}
\end{subfigure}
 \caption{A claim handling process as a motivating example.}
 \label{intro}
 \vspace{-20pt}
\end{figure}

For example, in the claim handling process shown in Fig.~\ref{intro_example}, a risk score assigned to the cases affects the handling procedure. On top, the business process explaining the whole process is shown without giving any clue that the process could be different for cases in different ranges of the risk score.  After creating an application, cases with a risk score higher than 10 are canceled. The remaining cases go through the check documents step and require an in-person interview, then there is an exclusive choice that decides whether the cases are required to submit more documents and have another interview, or they can skip these two steps if the risk score is lower than 3.  

Different dimensions for each case in the event log can be extracted using the available data in information systems and process analysis tools.
 The introduced approach in Fig.~\ref{ov_new} considers risk score as an example of a continuous dimension and has two main parts. First, the \textit{control flow change detection} is designed by using a sliding window shifting over the dimension range.
 Then, local peaks are used for \textit{segmentation and pairwise comparison}.

In this paper, we have the following goals:
1) Motivating the problem of finding process variants over the range of a continuous dimension, e.g., risk score or case duration.
2) Introducing a sliding window approach using the earth mover's distance to find the changes in control flow.
3) Case segmentation based on continuous dimensions and control flow, and pairwise comparison of the segments.
4) Testing the usability of the framework with a real-life case study.

The output of the framework is an automated segmentation of cases based on key dimensions influencing the control-flow behavior which can be used as input for other types of analysis, such as process comparison \cite{taymouri2021business}, outcome prediction \cite{TeinemaaDRM19}, or labeling the traces as desirable or undesirable in order to use them in process discovery algorithms working with desirable and undesirable traces~\cite{IMbi2023,ChesaniFGGLMMMT22}.

 \vspace{-10pt}
\section{Related Work}
% \vspace{-10pt}
Extracted event logs from information systems often consist of a variety of process executions containing deviating behaviors, manual interventions, infrequent patterns, process drift over time, and many other inconsistencies resulting in high complexity \cite{de2016general}. While filtering out infrequent behavior can mitigate data complexity \cite{Chapela-CampaML20}, it is important to note that infrequent behavior may sometimes include significant deviations that need further investigation. 
In \cite{leemans2020identifying}, the earth mover's distance is used to identify distinct cohorts of traces based on trace attributes, offering a framework that explores all trace attributes to uncover combinations leading to the most diverse control-flow behavior. 
Unlike this paper, we do not need the assumption that the dimensions are discretized in advance.

Trace clustering can also be considered as related work to our research. Different trace clustering approaches may use a control-flow perspective, other available attributes, or a combination of both \cite{clustering2015}. In \cite{BackS23}, different similarity measures to cluster the traces based on the control flow are introduced and compared. In \cite{ZelstC20}, an event log is divided into sub-event logs considering an attribute value, then hierarchical clustering is used to merge similar clusters. However, choosing a distinctive attribute is not straightforward. In addition, when considering continuous attributes it is even more challenging to set distinctive thresholds. 
In \cite{WeerdtBVB13}, the active trace clustering framework is introduced which takes process models explaining the clusters into account while doing trace clustering. The trace clustering methods cannot directly provide solutions for our problem, i.e., considering the performance perspective.

Considering time as an important dimension in event data analysis, the identification of concept drift is an interesting research question \cite{SatoFBS22}.
These methods usually use a feature space to characterize the control flow and use some techniques to pinpoint the changes. If a change is not observable using the selected feature space, these algorithms fail to detect it.
In \cite{YeshchenkoCMP22}, a concept drift detection algorithm is introduced which leverages declarative constraints to represent control-flow behavior. A multi-variate change point detection algorithm is implemented to find the changes over time. In \cite{MaaradjiDRO17}, another method is proposed which is based on the statistical tests applied to the distribution of partially ordered runs in two consecutive time windows. 
In \cite{brockhoff2020time}, the earth mover's distance is used to find drifts in the control flow. The mentioned approaches only focus on changes in control flow behavior over time and not on different continuous dimensions. Similar to the framework proposed in \cite{brockhoff2020time}, we employ a window-based approach using the earth mover's distance function to identify the changes across continuous features. Our framework works effectively without a large feature space, which might struggle to capture all potential changes in control flow. Additionally, it does not use statistical tests that often rely on some assumptions.

\vspace{-10pt}
\section{Preliminaries}
$\mathcal{B}(A)$ is the set of all multisets over some set $A$. Considering $B \in \mathcal{B}(A)$, $B(a)$ denotes the frequency of element $a \in A$. We write $x \in B$ if $B(x)>0$. 
The event log is an important concept in our work, therefore we formally introduce it.

\begin{definition}[Event Log]
Let $\mathcal{C}$, $\mathcal{A}$, and $\mathcal{T}$ be the universe of case identifiers, the universe of activities, and the universe of timestamps respectively. $e{=}(c,a,t)$ is a tuple representing an event where $\pi_{\mathcal{C}}(e){=}c \in \mathcal{C}$, $\pi_{\mathcal{A}}(e){=}a \in \mathcal{A}$, and $\pi_{\mathcal{T}}(e){=}t \in \mathcal{T}$. $\mathcal{E}{=}\mathcal{C} \times \mathcal{A} \times \mathcal{T}$ is the universe of events. A trace is a finite sequence of events $\sigma{=}\langle e_1, e_2, ..., e_{n} \rangle \in \mathcal{E}^{*}$ with size $n \in \mathbb{N}$ such that for each $1 \leq i < n$, $\pi_{\mathcal{C}}(e_i){=}\pi_{\mathcal{C}}(e_{i+1}) \wedge \pi_{\mathcal{T}}(e_i)\leq \pi_{\mathcal{T}}(e_{i+1})$. An event log $L$ is a set of traces such that each trace belongs to a different case.
% , $L=\{\sigma \vert c \in \mathcal{C}\}$. 
$\mathcal{L}$ is the universe of event logs. $\vert L \vert$ is the number of traces in event log $L$. In addition, we define $cf: \mathcal{L} \rightarrow \mathcal{B}(\mathcal{A}^*)$ as a function that extracts the control flow of $L$, i.e., the multiset of traces projected on activities such that 
$cf(L) {=} [\langle \pi_{\mathcal{A}}(e_1), ..., \pi_{\mathcal{A}}(e_n) \rangle|  \langle e_1, e_2, ..., e_{n} \rangle \in L]$.
\end{definition}

In addition to the control flow and time dimension, other dimensions could be assigned to cases from other sources of information. We use $\kappa_L: L \rightarrow \mathbb{R}$ to show a case-level indicator. Case-level indicators assign a numerical value $\kappa_L(\sigma)$ to each trace $\sigma {\in} L {\in} \mathcal{L}$.   If the context is clear, we drop $L$ from the notation $\kappa_L(\sigma)$.

\begin{definition}{(Ordering function)} 
Let $L \in \mathcal{L}$ be an event log and $\kappa$ be a case-level indicator that assigns a value to each case in $L$. We define $rank_{\kappa}(L) = \langle \sigma_1, \sigma_2, ..., \sigma_{\vert L \vert} \rangle$ such that $L{=}\{\sigma_1, \sigma_2, ... , \sigma_{\vert L \vert} \}$ and for $1 {\leq} i {<} j {\leq} \vert L \vert: \kappa(\sigma_i) {\leq} \kappa(\sigma_j)$. 
\end{definition}

 For example, $L=\{
\langle (c_1,a,13), (c_1,b,23) \rangle, 
\langle (c_2,a,14), (c_2,b,16) , (c_2,c,20) \rangle, $ $
\langle (c_3,a,17),(c_3,b,20),(c_3,c,35) \rangle \}$
is an event log with $cf(L) {=} [\langle a,b,c \rangle^{2}, \langle a,b \rangle^{1}]$. Consider $\kappa$ as a function that calculates the duration of cases, therefore, 
 $\kappa = \{ (\sigma_{c_1},10), $ $(\sigma_{c_2},6), (\sigma_{c_3},18)\}$ and
 $rank_{\kappa}(L) =\langle \sigma_{c_2}, \sigma_{c_1}, \sigma_{c_3} \rangle$.

\begin{definition}{(Stochastic Language)} 
Given the universe on activities $\mathcal{A}$, $f:\mathcal{A}^* {\rightarrow} [0,1]$ is a stochastic language iff $\sum_{s \in \mathcal{A}^*} f(s){=}1$. $\mathcal{F}$ is the universe of stochastic languages.
\end{definition}

\begin{definition}{(Stochastic Language of an Event Log)} 
Let $L \in \mathcal{L}$ be an event log. $stoch: \mathcal{L} \rightarrow \mathcal{F}$ is a function that extracts the stochastic language of an event log such that $stoch(L) = \{ (s,p) \vert s \in cf(L) \wedge p=\frac{cf(L)(s)}{\vert L \vert} \}$.
\end{definition}

The earth mover's distance calculates the distance between two stochastic languages. We use this distance to compare the control flow in two event logs.

\begin{definition}{(The earth mover's distance)}
Let $L \in \mathcal{L}$, and $L^{\prime} \in \mathcal{L}$ be two event logs, $\delta: \mathcal{A}^{*} \times \mathcal{A}^{*} \rightarrow [0,1] $ be a trace distance function, and $r: L \times L^{\prime} \rightarrow [0,1]$ be a function that indicates the movement of frequency between two event logs. $\mathcal{R}$ is the universe of all reallocation functions. The earth mover's distance between $L$ and $L^{'}$ is defined by
{\small
$EMD(L,L^{\prime})=\min\limits_{r \in \mathcal{R}} \sum_{s \in cf(L)} \sum_{t \in cf(L^{\prime})} r(s,t) . \delta(s,t)$} such that {\small
$\forall s \in cf(L): \sum_{t\in cf(L^{\prime})} r(s,t)={stoch(L)}(s)$}, {\small
$\forall t \in cf(L^{\prime}): \sum_{s\in cf(L)} r(s,t)={stoch(L^{\prime})}(t)$}, and {\small
$\forall s \in cf(L), \forall t \in cf(L^{\prime}): r(s,t) \geq 0$}.

\label{def:emd}
\end{definition}

We use the Levenshtein distance $\delta$ to calculate the distance between the traces. The calculation of the efficient reallocation amounts, i.e., $r$ in Def.~\ref{def:emd} is solved as a linear programming problem. For more details, we refer to~\cite{leemans2019earth}.

\vspace{-10pt}
\section{Process Variant Identification Framework}
As illustrated in Fig.~\ref{ov_new}, the framework proposed in this paper consists of two main parts, a control flow change detection over the range of a case-level indicator and a segmentation and comparison framework. Unlike concept drift frameworks which focus on time dimension, we identify changes in control flow across continuous dimensions. Our framework does not generate a large feature set that might overlook certain types of changes. Instead, it leverages the earth mover's distance to effectively capture the variability in control flow.

\subsection{Control Flow Change Detection}
\begin{figure}[htb]
\vspace{-10pt}
\begin{subfigure}{.50\textwidth}
\centering
\includegraphics[scale=0.50]{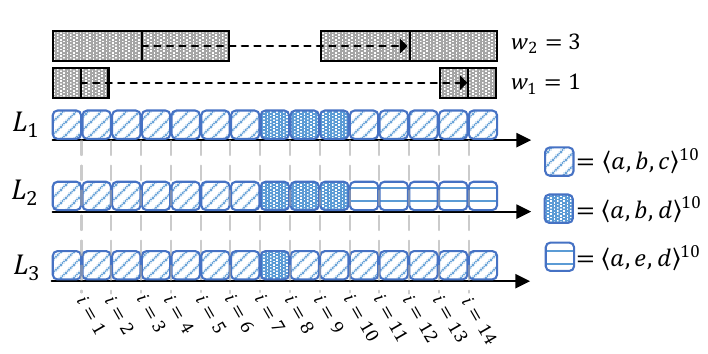}
\caption{Event logs $L_1$, $L_2$, and $L_3$ with different types of changes in the control flow.}
\label{window_shifting-exmple1}
\end{subfigure} \;
\begin{subfigure}{0.45\textwidth}
\centering
\includegraphics[scale=0.52]{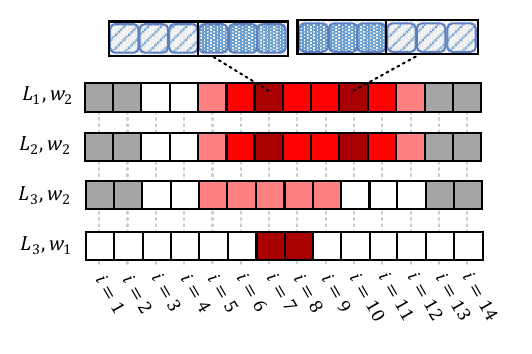}
\caption{Visualised $ldist$ value with moving the sliding window.}
\label{window_shifting-exmple2}
\end{subfigure}
 \caption{\footnotesize Some examples illustrating how the framework works considering three different event logs, $b=15$, and two time windows $w_1=1$ and $w_2=3$.}
 \label{window_shifting}
 % \vspace{-15pt}
\end{figure}

Our proposed method consists of a bucketing step and then moving a sliding window.
In Fig.~\ref{window_shifting-exmple1}, three different event logs, i.e., $L_{1}$, $L_{2}$, and $L_{3}$ are illustrated. Each of the event logs has 150 cases and the cases are ordered based on a case-level indicator. A bucketing strategy is used to divide the event logs into 15 buckets each containing 10 traces. The buckets are shown as patterned boxes such that different patterns show that the traces are different. In this example, two different window sizes $w_1{=}1$ and $w_2{=}3$ are used. Considering event log $L_1$ and window size $w_1$, we start from $i=1$ and each time we move the central point one unit to the right up to $i=14$ and compare $w_1$ bucket on the left-hand side of the central point to $w_1$ bucket on the right-hand side of it using the earth mover's distance. Comparative results for some combinations of the event logs and time windows are illustrated in Fig.~\ref{window_shifting-exmple2} using a color range proportional to the difference level. 
Next, we formally define buckets and sliding windows.

\begin{definition}{(Buckets)}
Let $L {\in} \mathcal{L}$ be an event log, $\kappa$ be a case-level indicator, $\sigma_i = rank_{\kappa}(L)(i)$, and $b \in \mathbb{N}^{[2,\vert L \vert]}$ be the number of buckets given by the user. For the sake of simplicity, we assume that $\vert L \vert$ is divisible by $b$. Then, $l{=}\frac{\vert L \vert}{b}$ is the number of cases in each bucket such that $B_i = \langle \sigma_{(i-1).l+1}, ..., \sigma_{i.l} \rangle$ for $1 \leq i \leq b$.
\end{definition}

\begin{definition}{(Left and Right Windows)} Let $L {\in} \mathcal{L}$ be an event log, $\kappa$ be a case-level indicator, $\sigma_i = rank_{\kappa}(L)(i)$, $b \in \mathbb{N}^{[2,\vert L \vert]}$ be the number of buckets given by the user, and $w \in \mathbb{N}^{[1,\frac{b}{2}]} $ be a window size parameter. Considering a window size $w$ and $i {\in} \{w,...,b-w \}$, we can create a left window
$W^{l}_{w,i}{=}\{\sigma {\in} B_j \vert i-w < j \leq i\}$
and a right window
$W^{r}_{w,i}{=}\{\sigma {\in} B_j \vert i < j \leq i+w\}$.
% \end{equation*}
\end{definition}

To perform a comparative analysis, we move the sliding window over the whole range of $\kappa$. We need at least $w$ buckets on the left window. Therefore, we start from $i{=}w$ and each time move the sliding window for one bucket. We repetitively continue until we reach a point where the number of buckets on the right window is equal to the window size, i.e., $i{=} b {-} w$.

\begin{definition}{(Local Distance Function)}
Let $L \in \mathcal{L}$ be an event log, $\kappa$ be a case-level indicator, $b \in \mathbb{N}^{[2,\vert L \vert]}$ be the number of buckets given by the user, and $w \in \mathbb{N}^{[1,\frac{b}{2}]} $ be a window size parameter. We define the local distance function $ldist_{L, \kappa, w, b}:\{w,...,b-w \} \rightarrow [0,1]$ such that {\small $ldist_{L, \kappa, w, b}(i) {= }EMD(W^{l}_{w,i},W^{r}_{w,i})$}. If the context is clear, we show $ldist_{L, \kappa, w, b}$ as $ldist$.
\end{definition}

In Fig.~\ref{window_shifting-exmple2}, some analysis is performed based on different event logs and different window sizes. Using the event log $L_{1}$ and the window size $w_2$, from $i{=}5$ the change in behavior is detectable by calculating the earth mover's distance between the left and right windows. The earth mover's distance value is visualized with colors, i.e., darker color shows a higher distance. After shifting the sliding window, the maximum difference between the left and right windows is observed at $i{=}7$ and $i{=}10$. A similar pattern in changing $ldist$ value is observed using $L_{2}$ and $w_2$. However, the introduced framework does not give us any clue whether the control flow behavior before $i{=}7$ and after $i{=}10$ are similar or different. Using $L_{3}$ and the smaller window $w_1$, it is observed that the change in behavior is only observed sharply at $i{=}7$ and $i{=}10$. Using the larger window $w_2$, the change in behavior started to affect $ldist$ from $i{=}5$ but with shifting the sliding window, it cannot exactly identify the point in which the behavior is changed. Therefore, the larger window size is more robust against noise but may miss some important change points.

\begin{figure}[tb]
\centering
\begin{subfigure}{0.65\textwidth}
% \centering
\includegraphics[scale=0.75]{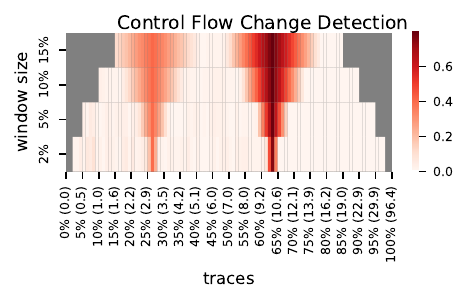}
% \vspace{-10pt}
\caption{Results of applying the introduced framework using the simulated event log with $b{=}100$ and $w \in \{ 2,5,10,15 \}$. }
\label{window_syn}
\end{subfigure} \;
\begin{subfigure}{0.27\textwidth}
% \centering
\includegraphics[scale=0.75]{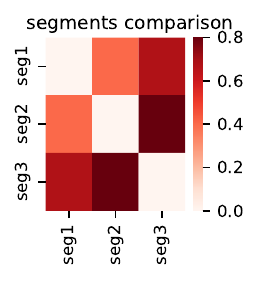}
% \vspace{-10pt}
\caption{Pairwise comparison between the segments.}
\label{heatmap_syn}
\end{subfigure}
\caption{Process variant identification framework for the claim handling process.}
\label{comp}
\vspace{-10pt}
\end{figure}

The running example introduced in Fig.~\ref{intro_example} is simulated using the CPN tools and the generated event log is used to explain how the framework works. This event log consists of 10,000 cases and 31 trace variants\footnote{\url{https://github.com/aliNorouzifar/process-variants-identification/blob/main/event\%20logs/test.xes}}. The cases are ordered based on their risk score value. In Fig.~\ref{window_syn}, the results using $b=100$, i.e., 1 \% of the cases in each bucket and different window sizes $w \in \{ 2,5,10,15\}$ are shown.  
The numbers in the parenthesis show the raw risk score values. The experiment is repeated for different window sizes. The colors show how high the earth mover's distance is between the left and right windows.

\subsection{Segmentation and Pairwise Comparison}
Based on the visualizations shown in Fig.~\ref{window_shifting}, the control flow behavior may change several times through the range of the indicator but with each change, we only know that $w$ buckets on the left are different from the $w$ buckets on the right. Therefore, we extend the proposed framework to make it more applicable to obtain global insights. The idea is to use the peaks in the $ldist$ values based on specific $b$ and $w$ parameters to perform a segmentation. Each time we observe a peak in $ldist$, we generate a segment consisting of the cases from the previous change point to the current change point. Then, it is possible to compare non-adjacent segments with the earth mover's distance measure.

\begin{definition}{(Change Point in Control Flow)}
Let $L \in \mathcal{L}$ be an event log, $\kappa$ be a case-level indicator, $b \in \mathbb{N}^{[2,\vert L \vert]}$ be the number of buckets given by the user, and $w \in \mathbb{N}^{[1,\frac{b}{2}]} $ be a window size parameter. $\theta \in [0,1]$ is a user-defined threshold to check whether a high value of $ldist$ is significant. $p \in [w,b-w]$ is a change point in the control flow behavior if $ldist(p) \geq \theta$, $ldist(p-1) \leq ldist(p)$ if $p \in (w,b-w]$, and $ldist(p+1) \leq ldist(p)$ if $p \in [w,b-w)$.
\end{definition}

\begin{definition}{(Segments)}
Let $P=\langle p_1,...,p_{\vert P \vert} \rangle$ be the ordered sequence of change points in $ldist$ function such that $\forall i,j \in \{1,..., \vert P \vert \}:p_i \leq p_j$ iff $i < j$. Considering $\vert P \vert$ as the number of peaks, we can generate $\vert P \vert +1$ segments which we refer to as $seg_1$ to $seg_{\vert P \vert+1}$ such that
 $seg_1= \{\sigma \in B_x \vert x \leq p_1\}$, $seg_i= \{\sigma \in B_x \vert p_{i-1} < x \leq p_{i}\}$ for $i \in [2,\vert P \vert]$, and $seg_{\vert P \vert + 1}= \{\sigma \in B_x \vert p_{\vert P \vert} < x\}$.
\end{definition}

 Considering $w=10$ in Fig.~\ref{window_syn} and $\theta=0.1$, $ldist$ has two peaks in $i=26$ ($risk \; score {=} 3$) and $i=63$ ($risk \; score = 10$). We can use heatmaps to compare the resulting segments pairwise to check if non-adjacent segments have similar control flows. Using the identified peaks, the whole event log is divided into three segments, and the segments are compared in Fig.~\ref{heatmap_syn}.

Next, we can hierarchically merge similar segments in case the control flow is not significantly different with regard to the user-defined difference threshold~$\theta$. 
% The algorithm for merging similar segments is explained in Algorithm~\ref{alg1}. 
We recursively merge two segments with the minimum $ldist$ if this minimum distance is lower than $\theta$, i.e., a user-defined threshold to check whether the distance is significant.

\section{Case Study}
The introduced framework is implemented and is publicly available\footnote{\url{https://github.com/aliNorouzifar/process-variants-identification}}. To assess the effectiveness of our framework, we conducted a comprehensive case study in collaboration with UWV, the employee insurance agency of the Netherlands responsible for executing social security in case of unemployment and disability. UWV provided real-life event data for this case study. We worked closely with experts from the agency, who provided invaluable guidance and insights. Their expertise enabled us to align our research with practical, real-world scenarios, ensuring that our algorithms are usable in real business contexts.

\subsection{UWV event log}
UWV has over 18000 employees and several branches all over the Netherlands. 
% There are various processes handled by this agency. 
One of the main processes of this governmental sector is investigated in this paper. 
The duration of the cases in this process has been of particular interest. UWV wants to know how the process is changed concerning the duration of the cases and whether we can find the change points in the duration. The duration of cases varies in the range of 1 day to 575 days.
The event log has 144,096 cases, 1,026 variants, 29 activities, and 1,316,128 events. 
Among all the cases, 5,449 cases (3.8\% of the cases) are rejected and 138,647 cases (96.2\% of the cases) are accepted.

\begin{figure}[tb]
\centering
\includegraphics[scale=0.1]{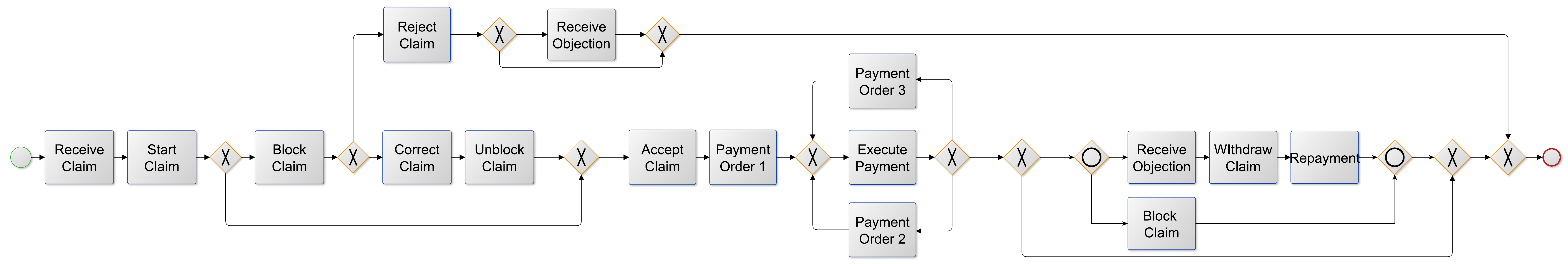}
\caption{\small  The normative BPMN model representing the investigated UWV claim handling process.}
\label{normative_model}
\end{figure}

The normative BPMN model shown in Fig.~\ref{normative_model} represents a claim-handling process at UWV. First, a claim from a customer is received. Then, either the claim is accepted or blocked. A blocked claim indicates either some information needs to be checked or corrected after which the claim is accepted, or the claim is blocked and then immediately rejected. After a rejected claim, an objection can be received by customers if they do not agree with the decision. The handling of this objection is out of the scope of this process model. After an accepted claim, which has received one or at most three payments, an objection can be received. This is due to customers who in hindsight find that they do not need the payments. A claim withdrawal process is started that results in repayment of the total sum of received benefits. In case the customer is still entitled to one or more payments, the \emph{Block Claim} activity is executed. This prevents any new payments from being automatically made. This normative model has an alignment fitness of 99.3\% with respect to the event log \cite{CarmonaDSW18}.

\vspace{-5pt}
% \vspace{-35pt}

% \vspace{-40pt}
\subsection{Process variant analysis using UWV event log}
\begin{figure}[b]
\vspace{-10pt}
% \centering
\begin{subfigure}{0.65\textwidth}
\centering
\includegraphics[scale=0.75]{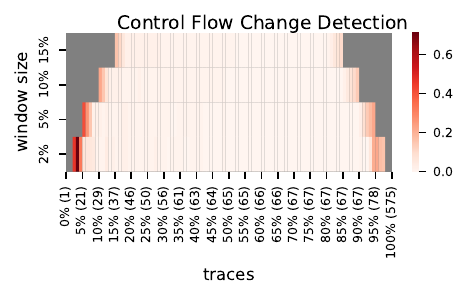}
\caption{\small Control flow change detection with $b{=}100$ and $w {\in} \{ 2, 5, 10, 15\}$.}
\label{window_UWV_comp}
\end{subfigure} \;
\begin{subfigure}{0.27\textwidth}
\centering
\includegraphics[scale=0.75]{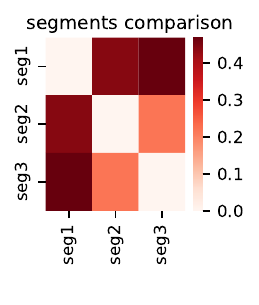}
\caption{\small Segmentation and pairwise comparison.}
\label{heatmap_uwv_comp}
\end{subfigure}
\caption{\small Applying the introduced framework to the complete UWV event log.}
\label{uwv_comp}
\vspace{-20pt}
\end{figure}

% \vspace{-20pt}
The first experiment is performed using the complete event log and the results are shown in Fig.~\ref{uwv_comp}. In this experiment, $b{=}100$ and the window sizes 2, 5, 10, and 15 are considered. The comparative results are illustrated in Fig.~\ref{window_UWV_comp} which shows that the cases with a very short duration or very long duration are different from other cases. Considering $\theta{=}0.1$, $w{=}5$, and $i$ referring to the central point of the sliding window, two peaks are observed at $i{=}3$ (17 days) and $i{=}95$ (78 days) which can be used to generate three segments, i.e., [0,17] days, [18,77] days and [78,575] days. In Fig.~\ref{heatmap_uwv_comp}, the three generated segments are compared to each other using the earth mover's distance.  

\begin{figure}[htb]
\begin{subfigure}{1\textwidth}
\centering
\includegraphics[scale=0.1]{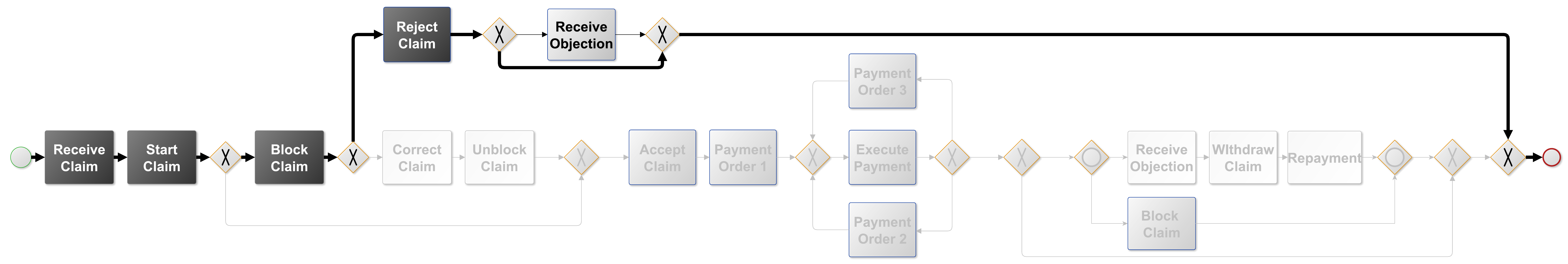}
\caption{\small Segment 1 in the complete event log with duration in the range [0,17] days.}
\label{all_seg1}
\end{subfigure}\\
\begin{subfigure}{1\textwidth}
\centering
\includegraphics[scale=0.1]{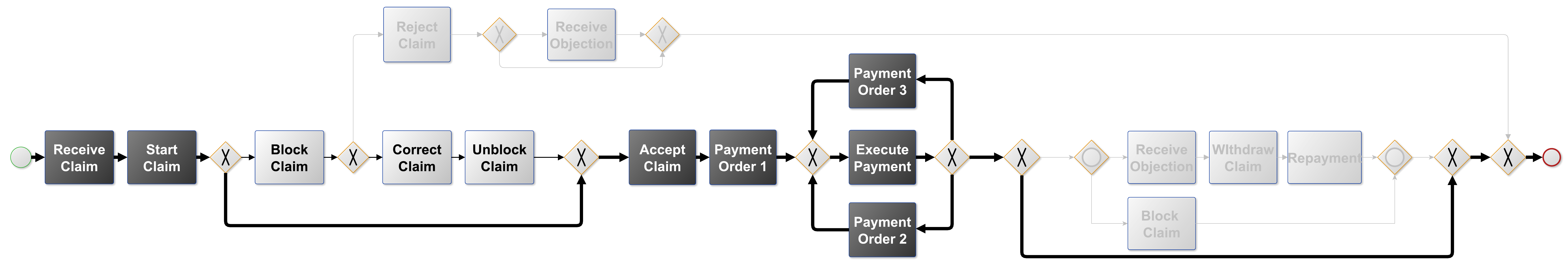}
\caption{\small Segment 2 in the complete event log with duration in the range [18,77] days.}
\label{all_seg2}
\end{subfigure}\\
\begin{subfigure}{1\textwidth}
\centering
\includegraphics[scale=0.1]{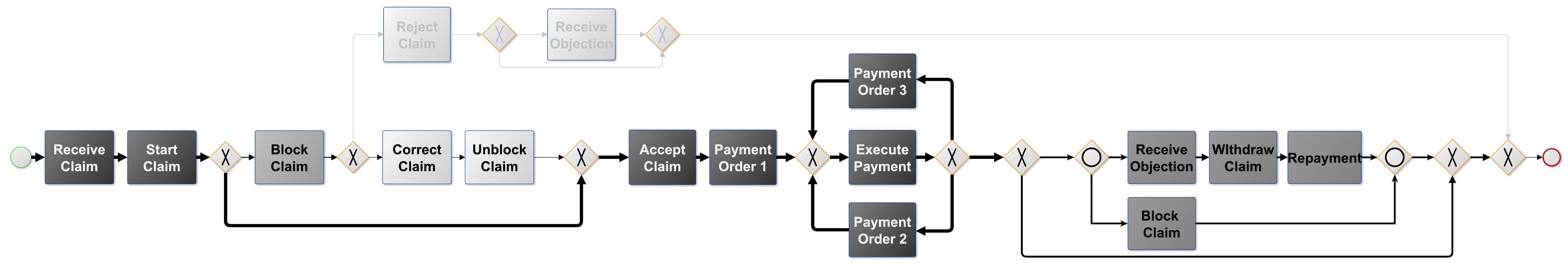}
\caption{\small Segment 3 in the complete event log with duration in the range [78,575] days.}
\label{all_seg3}
\end{subfigure}
\caption{\small The normative BPMN model is highlighted based on the frequency of the transitions in replaying the segments from the complete event log experiment.}
\label{uwv_process_all}
\vspace{-20pt}
\end{figure}

The number of cases in each segment and their alignment fitness with respect to the normative model is reported in Table.~\ref{fitness_table}. All the segments have a high fitness value implying that the observed behavior in the segments fits the normative model. According to the replay results, Fig.~\ref{uwv_process_all} shows the normative BPMN model where the parts of the model that cover each of the segments are highlighted. The colors are proportional to the frequencies and the transition labels are colored as gray if the frequency is lower than 5\%. For segment 1 in Fig.~\ref{all_seg1}, the activities \emph{Receive Claim}, \emph{Start Claim}, \emph{Block Claim}, \emph{Reject Claim} and \emph{Receive Objection} are highlighted. \emph{Segment 1} represents the claims that are rejected and optionally UWV receives an objection to the decision. \emph{Segment 2} shows claims that are accepted, i.e., the customer receives three payments and then the process finishes. Finally, \emph{segment 3} represents again accepted cases. However, after the payments an objection is received, and optionally a block is executed. For these claims, the customer 
withdraws the claim. The customer then pays the received sum back to UWV. 
Blocking is done to prevent new payments from being made while the claim withdrawal process has not finished yet.

\begin{table}[t]
\caption{\small Using alignment fitness as a conformance checking metric to check whether the event logs fit the process model or not.}
\label{fitness_table}
\centering
\begin{tabular}{|cl|cc|cc|}
\hline
\multicolumn{2}{|c|}{Event log}                                                                                     & \multicolumn{2}{c|}{Number of traces}                 & \multicolumn{2}{c|}{Alignment fitness}             \\ \hline
\multicolumn{1}{|c|}{\multirow{3}{*}{\begin{tabular}[c]{@{}c@{}}UWV\\ complete log\end{tabular}}}       & segment 1: [0,17] days & \multicolumn{1}{c|}{\multirow{3}{*}{144096}} & 4323   & \multicolumn{1}{c|}{\multirow{3}{*}{99.3\%}}  & 99.6\% \\ \cline{2-2} \cline{4-4} \cline{6-6} 
\multicolumn{1}{|c|}{}                                                                                  & segment 2: [18,77] days & \multicolumn{1}{c|}{}                        & 132572 & \multicolumn{1}{c|}{}                       & 99.7\% \\ \cline{2-2} \cline{4-4} \cline{6-6} 
\multicolumn{1}{|c|}{}                                                                                  & segment 3: [78,575] days & \multicolumn{1}{c|}{}                        & 7201   & \multicolumn{1}{c|}{}                       & 93.4\% \\ \hline
\multicolumn{1}{|c|}{\multirow{3}{*}{\begin{tabular}[c]{@{}c@{}}UWV\\ rejected claims\end{tabular}}}    & segment 1: [0,15] days & \multicolumn{1}{c|}{\multirow{3}{*}{5449}}   & 3888   & \multicolumn{1}{c|}{\multirow{3}{*}{98.4\%}} & 99.7\% \\ \cline{2-2} \cline{4-4} \cline{6-6} 
\multicolumn{1}{|c|}{}                                                                                  & segment 2: [16,83] days & \multicolumn{1}{c|}{}                        & 1350   & \multicolumn{1}{c|}{}                       & 97.3\% \\ \cline{2-2} \cline{4-4} \cline{6-6} 
\multicolumn{1}{|c|}{}                                                                                  & segment 3: [84,550] days & \multicolumn{1}{c|}{}                        & 211    & \multicolumn{1}{c|}{}                       & 81.7\% \\ \hline
\end{tabular}
\end{table}

\subsection{A deeper analysis considering rejected cases}
The extracted segments are highly correlated with the process outcome, i.e., the rejection or acceptance of a claim. However, the correlation between the extracted segments with specific duration is less clear. Another experiment using the 5,449 rejected cases is performed to get more understanding of the relation between duration ranges and specific process variants\footnote{The analysis of accepted cases is explained in the supplementary material \url{https://github.com/aliNorouzifar/process-variants-identification/blob/main/supplementary\%20material/supplementary\%20material.pdf}.}. The alignment fitness of the rejected cases is 98.04\% with respect to the normative model.

\begin{figure}[htb]
\vspace{-20pt}
\centering
\begin{subfigure}{0.65\textwidth}
\centering
\includegraphics[scale=0.75]{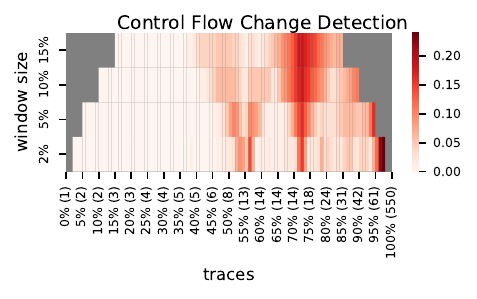}
\caption{\small Control flow change detection with $b{=}100$ and $w {\in} \{ 2, 5, 10, 15\}$.}
\label{window_UWV_comp0}
\end{subfigure} \;
\begin{subfigure}{0.27\textwidth}
\centering
\includegraphics[scale=0.75]{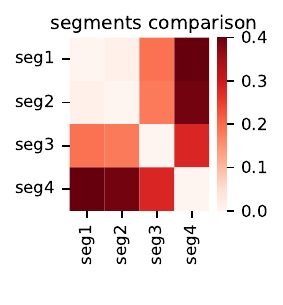}
\caption{\small Segmentation and pairwise comparison.}
\label{heatmap_uwv_comp0}
\end{subfigure}
\caption{\small Applying our framework to the rejected cases in the UWV event log.}
\label{uwv_comp0}
\vspace{-10pt}
\end{figure}

An overview of the results for the rejected cases is shown in Fig.~\ref{uwv_comp0}. Based on Fig.~\ref{window_UWV_comp0} considering $w=2$ and $\theta{=}0.1$, four segments are found, with duration periods of [0,13] days, [14,15] days, [16,83] days, and [84,550] days respectively. Segments 1 and 2 are merged into one segment, with a duration period of [0,15] days, since the distance between them is lower than $\theta{=}0.1$. This can be observed in the heatmap in Fig.~\ref{heatmap_uwv_comp0}. Segments 1 and 2 have very similar colors when compared to each other in comparison to the other segments.

\begin{figure}[htb]
\begin{subfigure}{1\textwidth}
\centering
\includegraphics[scale=0.1]{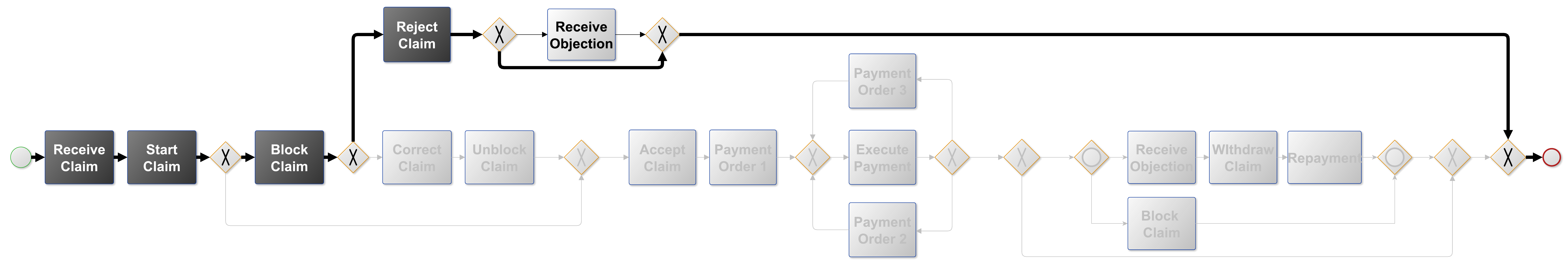}
\caption{\small Segment 1 in the rejected cases event log with duration in the range [0,15] days.}
\label{out0_seg1}
\end{subfigure}\\
\begin{subfigure}{1\textwidth}
\centering
\includegraphics[scale=0.1]{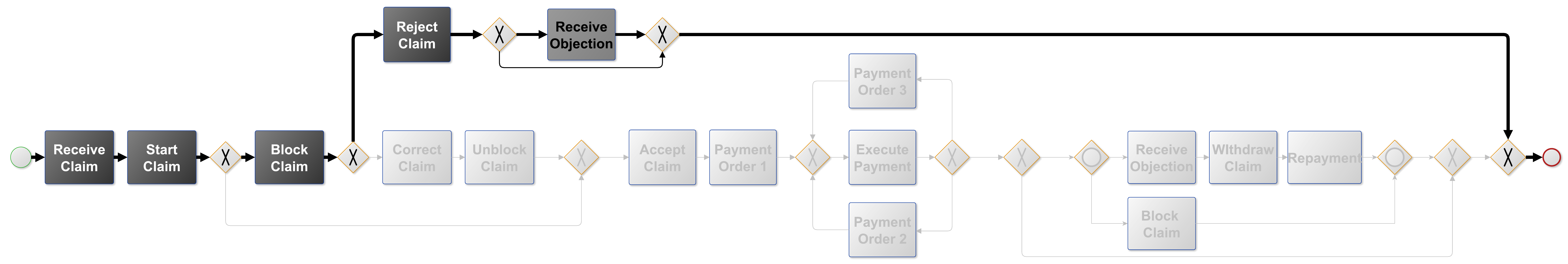}
\caption{\small Segment 2 in the rejected cases event log with duration in the range [16,83] days.}
\label{out0_seg2}
\end{subfigure}\\
\begin{subfigure}{1\textwidth}
\centering
\includegraphics[scale=0.1]{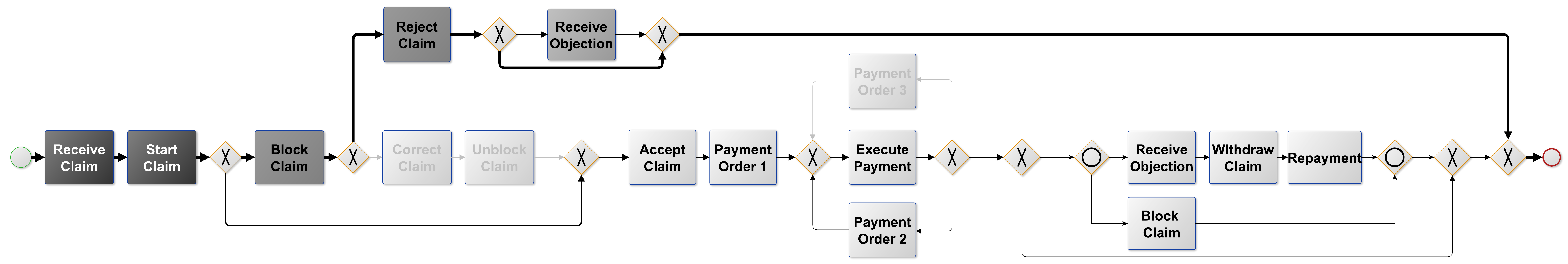}
\caption{\small Segment 3 in the rejected cases event log with duration in the range [84,550] days.}
\label{out0_seg3}
\end{subfigure}
\caption{\small The normative BPMN model highlighted based on the frequency of the transitions in replaying the segments from the rejected event log experiment.}
\label{uwv_process_out0}
\end{figure}

The number of cases in each segment and alignment fitness values of the identified segments are shown in Table~\ref{fitness_table}. These results show that segment 3 does not fit the normative model well with an alignment fitness value of 81.7\%. Fig.~\ref{uwv_process_out0} contains the highlighted normative model for each of the three segments. 
Fig.~\ref{out0_seg1} shows the first segment with a duration period of [0,15] days and contains cases that are rejected and no objection is received. This segment represents customers who file a claim even though they know that they most likely will not be entitled.
In the second segment, with a duration period [16,83] days, in Fig.~\ref{out0_seg2} cases are described that are rejected and some also have an objection. Finally, the third segment, with a duration period of [84,550] days, in Fig.~\ref{out0_seg3} consists of claims that are first accepted and end as rejection with a full repayment, or claims that are first rejected and are accepted after an objection is received and granted.

\section{Conclusion}
In this study, we delved into the concept of segmenting a continuous dimension concerning changes in control flow. While our framework primarily focuses on identifying segments with differing control flows, it may not directly imply the desirability or undesirability of individual cases. For instance, with a dimension like duration, cases within segments featuring either very short or very long duration could be associated with undesirability or efficiency. This approach holds promise for extension, particularly in identifying and labeling cases as undesirable, taking the broader process context into account. The interconnections between different dimensions may also play a role in the changes in behavior which is highly relevant for future investigations. Considering them may lead to some interesting analyses. For instance, the duration of cases could correlate with the workload of the process, potentially generating delayed yet normal cases during high workload periods. Considering only one dimension may not show whether cases are normal or problematic. 
The output of our framework can be used as input for various process mining tasks like process comparison, outcome prediction, and process discovery using desirable and undesirable cases.

\bibliographystyle{splncs04}
\bibliography{samplepaper}

\end{document}